\documentclass[preprint]{aastex63}

\shorttitle{HAWC and \textit{Fermi}-LAT detection of 2HWC J2006+341}
\shortauthors{Albert et al.}

\begin{document}
\title{HAWC and \textit{Fermi}-LAT Detection of Extended Emission from the Unidentified Source 2HWC J2006+341}

\correspondingauthor{Miguel Araya}
\email{miguel.araya@ucr.ac.cr}

\author{A.~Albert}
\address{Physics Division, Los Alamos National Laboratory, Los Alamos, NM, USA }

\author{R.~Alfaro}
\address{Instituto de F\'{i}sica, Universidad Nacional Aut\'onoma de M\'exico, Ciudad de M\'exico, Mexico }

\author{C.~Alvarez}
\address{Universidad Aut\'onoma de Chiapas, Tuxtla Guti\'errez, Chiapas, Mexico}

\author{J.R.~Angeles Camacho}
\address{Instituto de F\'{i}sica, Universidad Nacional Aut\'onoma de M\'exico, Ciudad de M\'exico, Mexico }

\author{M.~Araya}
\affiliation{Universidad de Costa Rica, San Jos\'e, Costa Rica}
\affiliation{Instituto Nacional de Astrof\'{i}sica, \'Optica y Electr\'onica, Puebla, Mexico }

\author{J.C.~Arteaga-Vel\'azquez}
\address{Universidad Michoacana de San Nicol\'as de Hidalgo, Morelia, Mexico }

\author{K.P.~Arunbabu}
\address{Instituto de Geof\'{i}sica, Universidad Nacional Aut\'onoma de M\'exico, Ciudad de M\'exico, Mexico }

\author{D.~Avila Rojas}
\address{Instituto de F\'{i}sica, Universidad Nacional Aut\'onoma de M\'exico, Ciudad de M\'exico, Mexico }

\author{H.A.~Ayala Solares}
\address{Department of Physics, Pennsylvania State University, University Park, PA, USA }

\author{V.~Baghmanyan}
\address{Institute of Nuclear Physics Polish Academy of Sciences, PL-31342 IFJ-PAN, Krakow, Poland }

\author{E.~Belmont-Moreno}
\address{Instituto de F\'{i}sica, Universidad Nacional Aut\'onoma de M\'exico, Ciudad de M\'exico, Mexico }

\author{C.~Brisbois}
\address{Department of Physics, University of Maryland, College Park, MD, USA }

\author{K.S.~Caballero-Mora}
\address{Universidad Aut\'onoma de Chiapas, Tuxtla Guti\'errez, Chiapas, Mexico}

\author{A.~Carrami\~nana}
\address{Instituto Nacional de Astrof\'{i}sica, \'Optica y Electr\'onica, Puebla, Mexico }

\author{S.~Casanova}
\address{Institute of Nuclear Physics Polish Academy of Sciences, PL-31342 IFJ-PAN, Krakow, Poland }

\author{U.~Cotti}
\address{Universidad Michoacana de San Nicol\'as de Hidalgo, Morelia, Mexico }

\author{E.~De la Fuente}
\address{Departamento de F\'{i}sica, Centro Universitario de Ciencias Exactas e Ingenier\'ias, Universidad de Guadalajara, Guadalajara, Mexico }

\author{C.~de Le\'on}
\address{Universidad Michoacana de San Nicol\'as de Hidalgo, Morelia, Mexico }

\author{R.~Diaz Hernandez}
\address{Instituto Nacional de Astrof\'{i}sica, \'Optica y Electr\'onica, Puebla, Mexico }

\author{B.L.~Dingus}
\address{Physics Division, Los Alamos National Laboratory, Los Alamos, NM, USA }

\author{M.A.~DuVernois}
\address{Department of Physics, University of Wisconsin-Madison, Madison, WI, USA }

\author{M.~Durocher}
\address{Physics Division, Los Alamos National Laboratory, Los Alamos, NM, USA }

\author{J.C.~D\'iaz-V\'elez}
\address{Departamento de F\'{i}sica, Centro Universitario de Ciencias Exactas e Ingenier\'ias, Universidad de Guadalajara, Guadalajara, Mexico }

\author{C.~Espinoza}
\address{Instituto de F\'{i}sica, Universidad Nacional Aut\'onoma de M\'exico, Ciudad de M\'exico, Mexico }

\author{H.~Fleischhack}
\address{Department of Physics, Michigan Technological University, Houghton, MI, USA }

\author{N.~Fraija}
\address{Instituto de Astronom\'{i}a, Universidad Nacional Aut\'onoma de M\'exico, Ciudad de M\'exico, Mexico }

\author{A.~Galv\'an-G\'amez}
\address{Instituto de Astronom\'{i}a, Universidad Nacional Aut\'onoma de M\'exico, Ciudad de M\'exico, Mexico }

\author{D.~Garcia}
\address{Instituto de F\'{i}sica, Universidad Nacional Aut\'onoma de M\'exico, Ciudad de M\'exico, Mexico }

\author{J.A.~Garc\'ia-Gonz\'alez}
\address{Instituto de F\'{i}sica, Universidad Nacional Aut\'onoma de M\'exico, Ciudad de M\'exico, Mexico }

\author{F.~Garfias}
\address{Instituto de Astronom\'{i}a, Universidad Nacional Aut\'onoma de M\'exico, Ciudad de M\'exico, Mexico }

\author{M.M.~Gonz\'alez}
\address{Instituto de Astronom\'{i}a, Universidad Nacional Aut\'onoma de M\'exico, Ciudad de M\'exico, Mexico }

\author{J.A.~Goodman}
\address{Department of Physics, University of Maryland, College Park, MD, USA }

\author{J.P.~Harding}
\address{Physics Division, Los Alamos National Laboratory, Los Alamos, NM, USA }

\author{B.~Hona}
\address{Department of Physics and Astronomy, University of Utah, Salt Lake City, UT, USA }

\author{D.~Huang}
\address{Department of Physics, Michigan Technological University, Houghton, MI, USA }

\author{F.~Hueyotl-Zahuantitla}
\address{Universidad Aut\'onoma de Chiapas, Tuxtla Guti\'errez, Chiapas, Mexico}

\author{P.~H\"untemeyer}
\address{Department of Physics, Michigan Technological University, Houghton, MI, USA }

\author{A.~Iriarte}
\address{Instituto de Astronom\'{i}a, Universidad Nacional Aut\'onoma de M\'exico, Ciudad de M\'exico, Mexico }

\author{A.~Jardin-Blicq}
\address{Max-Planck Institute for Nuclear Physics, 69117 Heidelberg, Germany}
\address{Department of Physics, Faculty of Science, Chulalongkorn University, 254 Phayathai Road,Pathumwan, Bangkok 10330, Thailand}
\address{National Astronomical Research Institute of Thailand (Public Organization), Don Kaeo, MaeRim, Chiang Mai 50180, Thailand}

\author{V.~Joshi}
\address{Erlangen Centre for Astroparticle Physics, Friedrich-Alexander-Universit\"{a}t Erlangen-N\"{u}rnberg, Erlangen, Germany}

\author{H.~Le\'on Vargas}
\address{Instituto de F\'{i}sica, Universidad Nacional Aut\'onoma de M\'exico, Ciudad de M\'exico, Mexico }

\author{J.T.~Linnemann}
\address{Department of Physics and Astronomy, Michigan State University, East Lansing, MI, USA }

\author{A.L.~Longinotti}
\address{Instituto Nacional de Astrof\'{i}sica, \'Optica y Electr\'onica, Puebla, Mexico }

\author{R.~L\'opez-Coto}
\address{INFN and Universita di Padova, via Marzolo 8, I-35131, Padova, Italy}

\author{G.~Luis-Raya}
\address{Universidad Polit\'ecnica de Pachuca, Pachuca, Hgo, Mexico }

\author{J.~Lundeen}
\address{Department of Physics and Astronomy, Michigan State University, East Lansing, MI, USA }

\author{K.~Malone}
\address{Physics Division, Los Alamos National Laboratory, Los Alamos, NM, USA }

\author{O.~Martinez}
\address{Facultad de Ciencias F\'{i}sico Matem\'aticas, Benem\'erita Universidad Aut\'onoma de Puebla, Puebla, Mexico }

\author{J.~Mart\'inez-Castro}
\address{Centro de Investigaci\'on en Computaci\'on, Instituto Polit\'ecnico Nacional, Ciudad de M\'exico, Mexico.}

\author{J.A.~Matthews}
\address{Dept of Physics and Astronomy, University of New Mexico, Albuquerque, NM, USA }

\author{P.~Miranda-Romagnoli}
\address{Universidad Aut\'onoma del Estado de Hidalgo, Pachuca, Mexico }

\author{E.~Moreno}
\address{Facultad de Ciencias F\'{i}sico Matem\'aticas, Benem\'erita Universidad Aut\'onoma de Puebla, Puebla, Mexico }

\author{M.~Mostaf\'a}
\address{Department of Physics, Pennsylvania State University, University Park, PA, USA }

\author{L.~Nellen}
\address{Instituto de Ciencias Nucleares, Universidad Nacional Aut\'onoma de M\'exico, Ciudad de M\'exico, Mexico }

\author{M.U.~Nisa}
\address{Department of Physics and Astronomy, Michigan State University, East Lansing, MI, USA }

\author{R.~Noriega-Papaqui}
\address{Universidad Aut\'onoma del Estado de Hidalgo, Pachuca, Mexico }

\author{N.~Omodei}
\address{Department of Physics, Stanford University: Stanford, CA 94305–4060, USA}

\author{A.~Peisker}
\address{Department of Physics and Astronomy, Michigan State University, East Lansing, MI, USA }

\author{E.G.~P\'erez-P\'erez}
\address{Universidad Polit\'ecnica de Pachuca, Pachuca, Hgo, Mexico }

\author{C.D.~Rho}
\address{University of Seoul, Seoul, Rep. of Korea}

\author{D.~Rosa-Gonz\'alez}
\address{Instituto Nacional de Astrof\'{i}sica, \'Optica y Electr\'onica, Puebla, Mexico }

\author{F.~Salesa Greus}
\address{Institute of Nuclear Physics Polish Academy of Sciences, PL-31342 IFJ-PAN, Krakow, Poland }
\address{Instituto de F\'isica Corpuscular, CSIC, Universitat de Val\`encia, E-46980, Paterna, Valencia, Spain}

\author{A.~Sandoval}
\address{Instituto de F\'{i}sica, Universidad Nacional Aut\'onoma de M\'exico, Ciudad de M\'exico, Mexico }

\author{M.~Schneider}
\address{Department of Physics, University of Maryland, College Park, MD, USA }

\author{R.W.~Springer}
\address{Department of Physics and Astronomy, University of Utah, Salt Lake City, UT, USA }

\author{K.~Tollefson}
\address{Department of Physics and Astronomy, Michigan State University, East Lansing, MI, USA }

\author{I.~Torres}
\address{Instituto Nacional de Astrof\'{i}sica, \'Optica y Electr\'onica, Puebla, Mexico }

\author{R.~Torres-Escobedo}
\address{Departamento de F\'{i}sica, Centro Universitario de Ciencias Exactas e Ingenier\'ias, Universidad de Guadalajara, Guadalajara, Mexico }

\author{F.~Ure\~na-Mena}
\address{Instituto Nacional de Astrof\'{i}sica, \'Optica y Electr\'onica, Puebla, Mexico }

\author{L.~Villase\~nor}
\address{Facultad de Ciencias F\'{i}sico Matemáticas, Benemérita Universidad Autónoma de Puebla, Puebla, Mexico }

\author{T.~Weisgarber}
\address{Department of Chemistry and Physics, California University of Pennsylvania, California, Pennsylvania, USA}

\author{E.~Willox}
\address{Department of Physics, University of Maryland, College Park, MD, USA }

\author{A.~Zepeda}
\address{Physics Department, Centro de Investigaci\'on y de Estudios Avanzados del IPN, Ciudad de M\'exico, Mexico }

\author{H.~Zhou}
\address{Tsung-Dao Lee Institute \& School of Physics and Astronomy, Shanghai Jiao Tong University, Shanghai, China}

\collaboration{74}{(HAWC Collaboration)}

\begin{abstract}
The discovery of the TeV point source 2HWC J2006+341 was reported in the second HAWC gamma-ray catalog. We present a follow-up study of this source here. The TeV emission is best described by an extended source with a soft spectrum. At GeV energies, an extended source is significantly detected in \textit{Fermi}-LAT data. The matching locations, sizes and spectra suggest that both gamma-ray detections correspond to the same source. Different scenarios for the origin of the emission are considered and we rule out an association to the pulsar PSR J2004+3429 due to extreme energetics required, if located at a distance of 10.8 kpc.
\end{abstract}


\section{Introduction} \label{sec:intro}
2HWC J2006+341 was discovered by the High Altitude Water Cherenkov observatory \citep[HAWC,][]{2017ApJ...843...40A} in the Cygnus Region of the Galaxy at the Galactic coordinates ${(l, b) = (71.33^\circ, 1.16^\circ)}$. No other TeV sources are known to exist nearby, and no supernova remnants (SNR) have been detected within one degree of 2HWC J2006+341 \citep{2014BASI...42...47G}. The pulsar PSR J2004+3429, located 0.4$\degr$ away from the position of 2HWC J2006+341, is the nearest pulsar found in the ATNF catalog \citep{2005AJ....129.1993M}. With a characteristic age of 18 kyr, the estimated distance and spin-down power of this pulsar are 10.8 kpc and $5.8\times 10^{35}$ erg s$^{-1}$, respectively \citep{2013MNRAS.435.2234B}.

In the GeV range, two point sources are found near 2HWC J2006+341 in the Fermi Large Area Telescope (LAT) fourth source catalog \citep[4FGL,][]{2020ApJS..247...33A}. 4FGL J2004.3+3339, $\sim 0.64\degr$ away, is associated to a binary system \citep[powering the nebula G70.7+1.2,][]{1992Natur.360..139K}. 4FGL J2005.8+3357, with a detection significance in the 4FGL catalog of 5.8$\sigma$, is located about $0.2\degr$ from 2HWC J2006+341 and has no association at other wavelengths.

After the discovery of 2HWC J2006+341 the MAGIC and \emph{Fermi}-LAT collaborations analyzed observations of the region to search for gamma rays \citep{2019MNRAS.485..356A}. Only upper limits were placed on the emission. Given the initially reported point source morphology for 2HWC J2006+341 by HAWC, no search for extended sources with radii above $0.2\degr$ was carried out by these authors.

This work expands our preliminary studies presented by \cite{2019ICRC...36..619A} with a more detailed analysis of HAWC and \emph{Fermi}-LAT data. Motivated by the possible presence of extended TeV emission in the 2HWC J2006+341 region in more recent and deeper HAWC maps, we searched for a corresponding signal in the GeV range with publicly available LAT data. The data from both observatories reveal emission with similar morphologies and consistent spectra, thus we believe that the GeV and TeV photons are produced by a single unidentified source and we discuss several scenarios for its nature.

\section{Data analysis}\label{sec:data}
\subsection{Fermi-\emph{LAT}}
The LAT onboard the \emph{Fermi} satellite is a converter/tracker telescope capable of detecting gamma rays in the energy range between 20 MeV and $\ga$1 TeV \citep{2009ApJ...697.1071A}. We analyzed Pass 8 LAT data from 2008 August to 2020 August with \emph{fermitools} version 1.2.23 and \emph{fermipy} version 0.19.0. Events with good quality in the SOURCE class were selected (with the parameter {\tt evclass=128}), including both back and front converting event types combined (with the parameter {\tt evtype=3}). The corresponding detector response {\tt P8R3\_SOURCE\_V2} was used. To reduce the gamma-ray background produced by the Earth's limb, only events having zenith angles lower than $90\degr$ were considered. Events were binned with a spatial scale of $0.05\degr$ per pixel and using ten bins per decade in energy for exposure calculation. The region of interest (ROI) analyzed had a radius of $15\degr$ and it was centered at the coordinates RA = $301.5 \degr$, Dec = $34.0 \degr$.

The energy range considered for spectral analysis was 1-500 GeV while the analysis of the morphology was carried out in the 5-500 GeV range to take advantage of the improved LAT resolution at higher energies. Sources found in the 4FGL catalog, located within $20\degr$ of the center of the ROI, were included in the analysis (162 in total). The diffuse Galactic emission and the isotropic emission (including the residual cosmic-ray background) were modeled with the standard files {\tt gll\_iem\_v07.fits} and {\tt iso\_P8R3\_SOURCE\_V2\_v1.txt}, respectively, provided by the LAT team\footnote{See \url{https://fermi.gsfc.nasa.gov/ssc/data/access/lat/BackgroundModels.html}}. The normalizations of the spectra of the 53 sources located within $10\degr$ of the centre of the ROI were left free to vary while their spectral shapes were fixed to their catalog values. All the spectral parameters of sources located farther away were fixed. The best-fit values of the free parameters and the optimization of the spectrum and morphology of the sources were obtained through the maximum likelihood technique \citep{1996ApJ...461..396M}, which maximizes the probability for the model to explain the data. This procedure allows for the estimation of the detection significance of a new source by calculating the test statistic (TS), defined as $-2\cdot$log$(\mathcal{L}_0/\mathcal{L})$, with $\mathcal{L}$ and $\mathcal{L}_0$ being the maximum likelihoods for models with and without the additional source, respectively. The weak source 4FGL J2005.8+3357, located in the region of 2HWC J2006+341 and having no association, was removed from the model in order to carry out a more detailed study of the emission.

In the first part of the analysis above 5 GeV, a search for extended emission in the region was carried out using two different morphological templates: an uniform disk and a symmetric Gaussian. A simple power-law spectral shape of the form $$\frac{dN}{dE} = N_0 \cdot \left(\frac{E}{E_0}\right)^{-\gamma},$$ where $E_0$ is a (constant) scale factor, was assumed (and later proved to be appropriate, see below). The location and size of the extended templates, as well as the spectral parameters, were fitted to maximize the likelihood. In order to compare models and since the models used are non-nested (it is not possible to go from one model to the other with a variation of the parameters) we applied the Akaike Information Criterion \citep[AIC,][]{1974ITAC...19..716A}. This is calculated as AIC = $2k - 2\ln(\mathcal{L})$ where $k$ is the number of parameters and $\mathcal{L}$ the maximum likelihood. Given a set of models, the one that better describes the data is the one that minimizes the AIC.

Once the best-fit morphology of the emission is determined, the spectra of the sources was obtained in the 1-500 GeV energy range. The spectral shape of the source of interest was changed for comparison and the new free parameters were fitted. The alternative spectral shape used was a log parabolic power law function (or log-parabola) of the form $$\frac{dN}{dE} = N_0 \cdot \left(\frac{E}{E_0}\right)^{-\alpha - \beta\, \ln(E/E_0)}.$$ The difference in TS values was used to indicate if there are significant deviations from a pure power law. In order to investigate the existence of additional sources in the region we carried out a systematic search for new point sources having a TS $>25$ and added them to the model. This was done before both the morphological and spectral analyses.

\subsection{\emph{HAWC}}
HAWC is a ground-based air shower detector array located at a latitude of $\sim19^{\circ}$ N in Mexico. Employing 300 water tanks of 7.3 m in diameter and 4.5 m in height each, it can scan about two thirds of the sky (8.4 sr, from $-26^{\circ}$ to $64^{\circ}$ declination) detecting very-high-energy gamma rays with energies from hundreds of GeV to $> 100$ TeV and a $>95\%$ duty cycle \cite{2017ApJ...843...40A,2017ApJ...843...39A,2019ApJ...881..134A}. The angular resolution depends on the source location in the sky and the energy of the events, and varies from $1^{\circ}$ to $0.2^{\circ}$ \citep{2017ApJ...843...39A}, which is similar to the resolution of the LAT \citep{2020ApJS..247...33A}. The data used, with a livetime of 1038.7 days, were binned using the fraction of detectors triggered by an air shower, and the ground parameter \citep[as defined in][]{2019ApJ...881..134A} was used to estimate the energy of the events. Recently optimized gamma/hadron separation cuts and reconstruction were used which were found to improve the energy and angular resolutions with respect to previous analyses \citep{2019ApJ...881..134A}. The Multi-Mission Maximum Likelihood framework\footnote{\url{www.github.com/threeML/}} \citep[3ML,][]{threeML} was used to perform the likelihood fits and determine the morphology and spectrum of the source. The HAWC accelerated likelihood\footnote{\url{www.github.com/threeML/hawc_hal}} (HAL) plugin was used.

\begin{figure}
\begin{center}
\includegraphics[width=14cm]{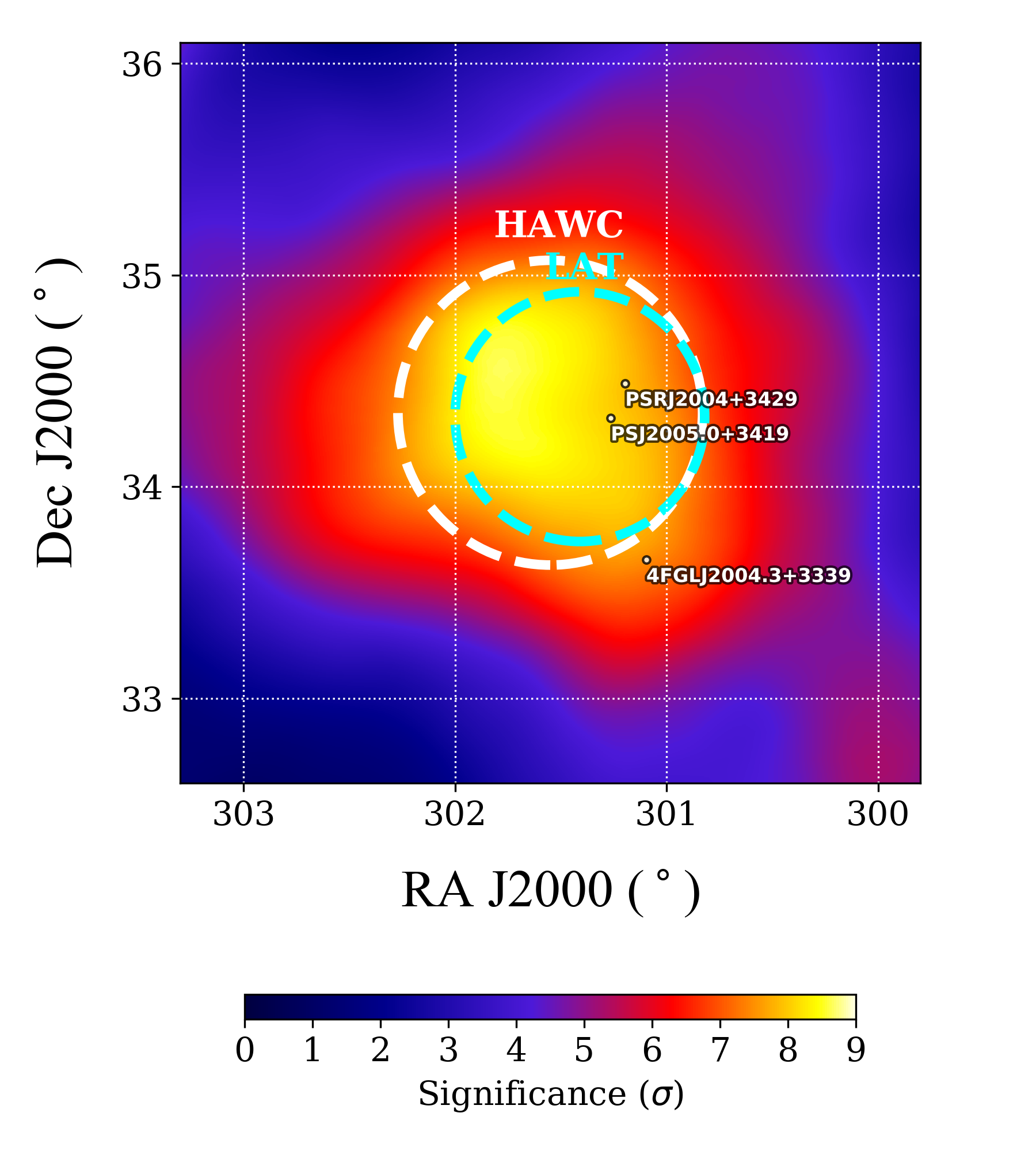}
\caption{HAWC significance map of the 2HWC J2006+341 region obtained using a 1$\degr$-extended source hypothesis with a simple power-law spectrum. The dashed lines mark the extension of the best-fit Gaussian templates ($\sigma$) found with the HAWC (white) and LAT (cyan) data. The locations of the 4FGL sources in the region are indicated as well as the position of the pulsar PSR J2004+3429 and the new point source found in this work, PS J2005.0+3419.}
\label{fig1:map}
\end{center}
\end{figure}

The ROI analyzed was a circular region with a radius of $7^{\circ}$ around the cataloged position of 2HWC J2006+341 \citep{2017ApJ...843...40A}. Besides the source of interest, the model contained the pulsar wind nebula (PWN) 2HWC J2019+367, located $\sim 3.6^\circ$ from the center of the ROI, and a background template of uniform brightness covering the entire ROI to account for emission from any possible unresolved sources. Changing the model of the gamma-ray background in the region would affect the parameters of the source of interest, and thus we considered an additional source of systematic uncertainty associated to this model. We calculated the deviations of the spectral parameters obtained when replacing the background template with two alternative models and optimizing the likelihood. In one alternative no gamma-ray background was included and for the other one we used a 2D Gaussian background centered at the Galactic plane. We added the differences between the resulting parameter values and the original ones in quadrature to the other systematic errors. Similarly to the analysis of LAT data, maximum likelihood fits were done using a point source hypothesis as well as the uniform disk and symmetric Gaussian templates, assuming a simple power-law spectrum for the emission. Once the best-fit morphology is found, a fit using a log-parabolic spectrum was done to search for deviations from the simple power-law (spectral curvature). Fig. \ref{fig1:map} shows a significance map of the 2HWC J2006+341 region calculated with the maximum likelihood ratio and an extended source with a power-law spectrum with an index $\gamma = 2.7$, which results from the initial fit. The results of the morphological analysis are also shown in the figure.

\section{Results}
In the LAT data no point sources having a TS$>25$ were found in the region of 2HWC J2006+341 above 5 GeV. However, significant emission was detected using extended templates which gave TS$> 40$ for both the disk and Gaussian above 5 GeV. The likelihood ratio between the best-fit extended model and the best-fit point source model was such that TS$_{ext}>20$ in both cases. Similarly, in HAWC data the extended templates result in significant detections. Table \ref{table1:morph} gives the fitted extensions, $\Delta$AIC and TS values of the fits for each analysis. Here, ${\Delta\mbox{AIC} = \mbox{AIC}_m - \mbox{AIC}_{min}}$ is the difference in AIC between each model $m$ and the one that minimizes the AIC ($\Delta$AIC = 0 for the best available model). In the analysis of HAWC data the Gaussian template provided the best description compared to the others. The fit with a Gaussian morphology also produced smaller residuals in HAWC data. For these reasons this template was chosen to represent the emission in the region of 2HWC J2006+341 at TeV energies. The locations of the Gaussian centres that maximized their respective likelihoods (with their respective $1\sigma$-level statistical uncertainties) are given by RA = $301.55 \pm 0.18^{\circ}$, Dec = $34.35 \pm 0.16^{\circ}$ in the HAWC analysis, and RA = $301.41\pm 0.15\,^{\circ}$, Dec = $34.33\pm0.16\,^{\circ}$ in the fits to the LAT data. The locations are compatible within uncertainties.

In the analysis of LAT data both extended morphologies tested represented a similar improvement with respect to the null hypothesis. The Gaussian template was chosen for the rest of the analysis for a better direct comparison with the HAWC results. The effect of changing the morphological model was considered a source of systematic uncertainty. The other source of systematic uncertainty in the LAT data was estimated by propagating the uncertainty in the effective area onto the spectral index and flux. A set of bracketing response functions were used for this purpose as recommended by \cite{2012ApJS..203....4A}. As a consistency check, we repeated the entire analysis after replacing the diffusion model with a different version, given by the file gll\_iem\_v06.fits, and obtained similar results. The source extension was also confirmed with TS$_{ext}>28$ after artificially varying the best-fit value of the diffuse Galactic normalization by $\pm6$\%, an estimate of the systematic uncertainties in the diffuse emission model used in the past by the LAT team \citep[e.g.,][]{Abdo2010}. Systematic uncertainties in the HAWC data result from uncertainties in the modeling of the detector. For the spectral parameters measured by HAWC, these uncertainties were estimated as explained in \cite{2019ApJ...881..134A}. New fits were done with alternative instrument response files changing key parameters used to model the instrument. The differences between the resulting spectral parameters and the nominal values were used to estimate the systematic uncertainties.

\begin{table}
\caption{Results of the independent morphological analyses of the \emph{Fermi}-LAT and HAWC data sets.}
\label{table1:morph}
\begin{center}
\begin{tabular*}{\textwidth}{@{\extracolsep{\fill} } clcccccc}
\hline
\hline
\textbf{Data set} & \textbf{Spatial model} & \textbf{Fitted size$^{a}$ ($\degr$)} & \textbf{RA ($\degr$)} & \textbf{Dec ($\degr$)} & \textbf{$\Delta$AIC}  & \textbf{TS}$^{b}$\\
\hline
\emph{Fermi}-LAT & & & & \\

& Disk & $0.85_{-0.05}^{+0.04}$ & $301.50\pm 0.06$ & $34.47 \pm 0.08 $ & 0 & 44.6\\

& Gaussian & $0.59_{-0.11} ^{+0.10}$ & $301.41\pm 0.15$ & $34.33\pm 16$ & 3.6 & 41.0\\

\hline
HAWC & & & & & & \\

& Point source & - & $301.1\pm 0.07$ & $34.16 \pm 0.05$ &  23.1 & 24.0\\

& Disk & $0.88\pm 0.10$ & $301.6 \pm 0.15$ & $34.5 \pm 0.15$ & 7.2 & 41.9\\ 

& Gaussian & $0.72\pm 0.14$ & $301.55 \pm 0.18$ & $34.35 \pm 0.16$ & 0 & 49.0\\
\hline
\end{tabular*}\\
\textsuperscript{$a$}\footnotesize{Radius for the disk and sigma for the Gaussian.}\\
\textsuperscript{$b$}\footnotesize{The LAT TS values were obtained with events above 5 GeV.}\\
\end{center}
\end{table}

\subsection{Spectrum of the HAWC source}
Using the corresponding (Gaussian) templates the spectra of the sources were obtained. An improvement in the fit quality was obtained in the HAWC data when fitting a log-parabola with respect to a simple power-law at the $\sim 4\sigma$ level. The corresponding TS values are 65.8 and 49.0 for the log-parabola and the simple power-law, respectively. The resulting best-fit values of the parameters for the HAWC source are ${N_0 = (7.6_{-1.5} ^{+1.8}(stat)_{-1.4} ^{+2.1}(sys))\times 10^{-14}\, \mbox{TeV}^{-1} \,\mbox{cm}^{-2} \,\mbox{s}^{-1}}$, $\alpha = 3.1 \pm 0.5(stat)_{-0.3}^{+0.4}(sys)$ and $\beta = 1.0 \pm 0.6(stat)_{-0.6}^{+1.0}(sys)$. The value of the scale factor was fixed to $E_0 = 7$ TeV following \cite{2019ApJ...881..134A}. With the morphological and spectral parameters found we fit a sharp cutoff at low and high energy \citep[see][]{2017ApJ...843...39A}. We found the highest value at low energies and the lowest value at high energies that are inconsistent with the observations at the 1$\sigma$ level. In HAWC data the resulting energy range where the source was significantly detected is 2--13 TeV.

\subsection{Spectrum and significance of the LAT source in the 1-500 GeV range}
Above 1 GeV the search for point sources returned one source in the region of 2HWC J2006+341, which we labeled PS J2005.0+3419, located at the coordinates RA = $301.26 \pm 0.02\degr$, Dec = $34.32 \pm 0.03 \degr$. A new fit including the extended source found above and this point source results in the TS values of 62.9 and 33.6, respectively. PS J2005.0+3419 shows a soft spectrum with an index of $2.9\pm0.2$. No significant spectral curvature was seen for the extended source. The TS values of the fits with a simple power-law and a log-parabola are TS$_{pl}$ = 62.9 and TS$_{logpar}$ = 63.6, respectively. The overall detection significance of the extended LAT source above 1 GeV is thus $\sim7.9\sigma$. The simple power-law fit gave a spectral index of ${1.85\pm 0.10(stat) \pm 0.03(sys)}$ and an integrated flux in the 1-500 GeV range of ${(1.8\pm 0.4(stat) \pm 0.5(sys))\times 10^{-9}\, \mbox{cm}^{-2}\, \mbox{s}^{-1}}$. 

The inclusion of PS J2005.0+3419 in the model resulted in a fully compatible spectrum for the extended source with respect to the results obtained above 5 GeV where PS J2005.0+3419 was not detected. A model containing both the extended source and PS J2005.0+3419 is also a better description of the emission above 1 GeV, with $\mbox{AIC}_{\mbox{ext}} - \mbox{AIC}_{\mbox{ext+ps}}\sim 20$.

\subsection{GeV--TeV spectral energy distribution}
\begin{figure}
\begin{center}
\includegraphics[width=\textwidth]{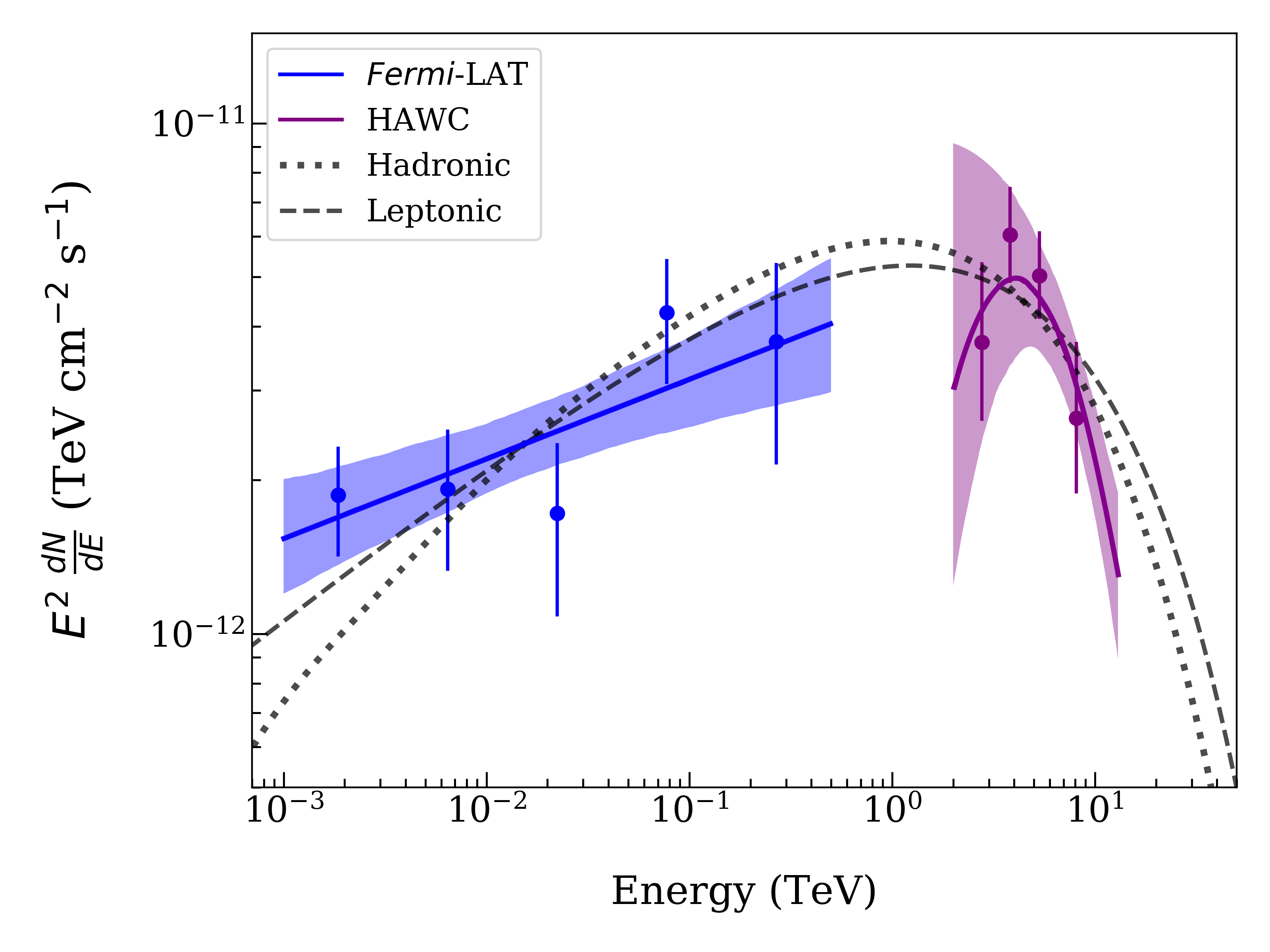}
\caption{SED of 2HWC J2006+341 resulting from the analysis of the HAWC and LAT data sets described in Section \ref{sec:data}. The blue solid line represents the best-fit LAT power-law spectrum while the purple line is the best-fit log-parabola from the HAWC analysis. Each shaded band represents the corresponding propagated $1\sigma$-statistical uncertainty. The dashed and dotted lines are the models described in Section \ref{discussion}.}
\label{fig2:SED}
\end{center}
\end{figure}

Since the morphologies of the LAT and HAWC sources are consistent with each other we attribute the emission to a single new extended gamma-ray source. Fig. \ref{fig2:SED} shows the spectral energy distributions (SED) obtained from both data sets and the models discussed in Section \ref{discussion}.

\section{Discussion}\label{discussion}
The analysis of the LAT data shows that there is an extended source of GeV emission in the region of 2HWC J2006+341. The size and location of this source are consistent with those of the TeV emission revealed in the detailed analysis of the HAWC data presented here. This result and the fact that the SEDs measured by both instruments show comparable fluxes, as can be seen in Fig. \ref{fig2:SED}, are  evidence that the particles responsible for the GeV and TeV gamma rays come from the same source. Furthermore, the overall TeV spectrum is not an extrapolation of the power-law spectrum seen with the LAT, which is harder, indicating the presence of spectral softening.

Regarding the origin of the gamma rays, both an SNR and a PWN could in principle be responsible, as they typically produce extended gamma-ray emission \citep[e.g.,][]{2016ApJS..224....8A,2018A&A...612A...1H,2020ApJS..247...33A}. In order to account for the SED seen in Fig. \ref{fig2:SED}, simple emission mechanisms known to exist within these objects can be explored to get an idea of the energetics in the particles producing the radiation. For this purpose, fits to the data points shown in Fig. \ref{fig2:SED} were done with the Markov Chain Monte Carlo fitting routines of {\tt naima}, a package for the calculation of non-thermal emission from relativistic particles \citep{naima}. Using a particle energy distribution that is a power-law with an exponential cutoff, the data were fit to fluxes resulting from both leptonic and hadronic scenarios. In the leptonic scenario the particles interact with the cosmic microwave background (CMB) bringing these photons to gamma-ray energies through inverse Compton (IC) scattering \citep[for details on the implementation of the calculation, see][]{2014ApJ...783..100K}. In the hadronic scenario, relativistic protons produce gamma rays through collisions with ambient protons \citep[see][]{2014PhRvD..90l3014K}. The best-fit models obtained in both scenarios are plotted in Fig. \ref{fig2:SED}. The resulting AIC values are 13.2 (for the leptonic model) and 15.4 (for the hadronic scenario).

PWNe are believed to produce gamma rays mainly from electrons and positrons interacting with soft background light, and therefore because these particles cool off efficiently as they propagate away from the source the size of PWNe resulting from IC scattering usually decreases with increasing energy. This prediction is consistent with observations of HESS J1825-137 and Geminga \citep{2006A&A...460..365A,2019PhRvD.100l3015D,2019ICRC...36..595P}. However, the results of the morphological analysis of 2HWC J2006+341 presented here reveal extensions at GeV and TeV energies that are consistent within uncertainties. This could disfavor the PWN origin for the emission and deeper observations will be required to confirm this. In the leptonic IC-CMB scenario, the resulting spectral index and cutoff energy of the particle distribution, as well as the total energy in the particles above 1 GeV (and their corresponding $1\sigma$ statistical uncertainties) are $2.38^{+0.10} _{-0.19}$, $(42^{+19}_{-15})$ TeV and $(8.8^{+6.6}_{-5.1})\times 10^{47}\,(d/\mbox{1 kpc})^2$ erg, respectively. Here, $d$ is the source distance. As noted in Section \ref{sec:intro} the only pulsar known near the gamma-ray source is PSR J2004+3429, with an estimated distance of 10.8 kpc. Assuming the same distance to 2HWC J2006+341 the required energy in the leptons would be in the range $10^{49}-10^{50}$ erg, considering the uncertainty. A rough estimate of the energy injected by the pulsar can be obtained by the product of the spin-down power with its characteristic age, which yields $\sim4\times 10^{47}$ erg. Assuming a braking index for the pulsar of 2, rather than the default value of 3, could change this estimate by a factor of a few \citep{2006ARA&A..44...17G}, but the energy in the particles would be two orders of magnitude above this value, making it difficult to reconcile the energetics. If additional photon fields are considered in the calculation of the IC fluxes the required energy in the leptons is still very large. For temperatures and densities of far-infrared dust emission and near-infrared stellar emission seen in the Solar System the total required energy decreases by a factor of $\sim 3$ with respect to the IC-CMB scenario, which is not enough.

Assuming the same distance of 10.8 kpc as PSR J2004+3429 for the gamma-ray source, the luminosity in the 2--13 TeV range would be $\sim 2\times10^{35}$ erg s$^{-1}$, or about 7 times the luminosity of the Crab Nebula in the same energy range. This TeV luminosity for 2HWC J2006+341 and the characteristic age of pulsar PSR J2004+3429 would be similar to the corresponding values for the firmly identified PWN HESS J1825-137 and its associated pulsar. However, the spin-down power of PSR J2004+3429 is almost five times lower than that of the pulsar associated to HESS J1825-137. This would make 2HWC J2006+341 fall outside of the trends seen for other PWNe regarding their TeV luminosities \citep{2018A&A...612A...2H}. Furthermore, the feature that would make 2HWC J2006+341 a very unusual PWN would be its large physical size. An extension of $\sim 1.7^{\circ}$ in the sky corresponds to a size of more than 300 pc for a distance to PSR J2004+3429 of 10.8 kpc. It then becomes difficult to propose PSR J2004+3429 as the source of the gamma-ray emission from 2HWC J2006+341. Although it is possible in principle for 2HWC J2006+341 to be a PWN produced by an unknown pulsar located at a closer distance, the similar sizes at GeV and TeV energies found here could be in conflict with the PWN scenario. Deeper observations with more statistics and a detailed study of the environment at the location of 2HWC J2006+341 are necessary to confirm or reject this hypothesis.

The discovery by HAWC of extended TeV emission around the Geminga and PSR B0656+14 pulsars \citep{2017Sci...358..911A} points to the possible existence of a population of middle-aged pulsars producing ``TeV halos'' in the surrounding interstellar medium \citep[e.g.,][]{2017PhRvD..96j3016L}. The possibility that a previously undetected pulsar is responsible for the gamma rays from 2HWC J2006+341 cannot be discarded. However at GeV energies the likely counterpart of the Geminga halo is much more extended than the TeV emission \citep{2019PhRvD.100l3015D}, which is not seen for 2HWC J2006+341.

Another possibility for the origin of 2HWC J2006+341 is the SNR shell scenario. SNRs are mainly seen in radio observations. However, inspection of radio images from the 1.4 GHz NRAO VLA Sky Survey \citep[NVSS,][]{1998AJ....115.1693C} and the Westerbork Northern Sky Survey \citep[WENSS,][]{1997A&AS..124..259R} show no obvious hints of emission. The known SNR that is closest in the sky to 2HWC J2006+341 is $\sim1.8^{\circ}$ away \citep[G69.7+1.0,][]{2019JApA...40...36G}. The lack of a known counterpart at lower wavelengths is not necessarily evidence against the SNR scenario. There are several SNR shell candidates such as HESS J1614-518 that have only been detected at TeV energies \citep{2018A&A...612A...8H}. Using the same simple leptonic scenario described above, the total energy in the particles is reasonable for a large range of possible distances to the source. The typical kinetic energy in the shocks of SNRs that is available for accelerating particles is of the order of $E_{SN}=10^{51}$ erg. The required total energy in the leptons above 1 GeV, $\sim 9\times 10^{47}\,(d/\mbox{1 kpc})^2$ erg, is not problematic for a large range of possible distances to the source. Given the extension of 2HWC J2006+341 and that the typical diameters of SNRs are of tens of pc, the maximum distance to the source would likely be several kpc. For a source distance of 1 kpc, for instance, the diameter of the gamma-ray emission region would be $~30$ pc and the energy content in the leptons (above a particle energy of 1 GeV) $10^{47}-10^{48}$ erg, under the simple one-zone model. This energy is comparable to the energy content in leptons estimated, for example, in the SNR RCW 86 \citep{2018A&A...612A...4H}.
 
The gamma rays could also be caused by hadronic interactions between energetic cosmic rays, accelerated in an SNR, and ambient matter. The fit to the data yielded a particle spectral index of $1.69^{+0.06} _{-0.9}$ and a cutoff energy of $58^{+14}_{-16}$ TeV. The required total energy in the hadrons, in terms of the distance to the source and the number density of the target material ($n$) is $(2.2\pm 0.2)\times 10^{49} \,(d/\mbox{1 kpc})^2$ (1 cm$^{-3}/n)$ erg. For a range of plausible distances and typical gas densities the required total energy in the particles could be consistent with theory and observations \citep[$\sim 0.03 - 0.3 \, \, E_{SN}$, e.g.,][]{1994A&A...287..959D,2013Sci...339..807A}. However, the particle distribution derived from the data is harder than that predicted in the test-particle diffusive shock acceleration theory \citep[see, e.g.,][]{1978MNRAS.182..147B}, and considerably harder than inferred for typical hadronic SNRs \citep[e.g.,][]{2013Sci...339..807A}. Thus the leptonic SNR scenario might be a more plausible explanation for the origin of the emission.

Deeper observations with HAWC and more detailed studies looking for counterparts at other wavelengths are necessary to understand the nature of the source 2HWC J2006+341. The work described here is a first step in this direction. For example, observations of this source in the X-rays could help constrain the maximum electron energies while lower energy gamma-ray observations could probe for the characteristic pion bump of the hadronic scenario. Follow-up observations of the GeV point source PS J2005.0+3419 could also reveal its nature and any possible association to the GeV--TeV emission. Radio observations could find a previously undetected SNR. The matching morphologies at GeV and TeV energies make 2HWC J2006+341 consistent with an SNR shell with no clear counterparts at lower energies, perhaps similar to the TeV sources HESS J1614-518 and HESS J1912+101, or to the extended GeV source G350.6-4.7 \citep[with a spectral index of $\sim1.7$ and no known counterpart at lower energies,][]{2018MNRAS.474..102A} and other similar objects \citep{2018ApJS..237...32A}, having a similar spectrum at GeV energies. Consistent models for a possible population of such objects have yet to be explored.

\acknowledgments

We thank the anonymous referee for helpful suggestions that improved this work. We acknowledge the support from: the US National Science Foundation (NSF); the US Department of Energy Office of High-Energy Physics; the Laboratory Directed Research and Development (LDRD) program of Los Alamos National Laboratory; Consejo Nacional de Ciencia y Tecnolog\'ia (CONACyT), M\'exico, grants 271051, 232656, 260378, 179588, 254964, 258865, 243290, 132197, A1-S-46288, A1-S-22784, c\'atedras 873, 1563, 341, 323, Red HAWC, M\'exico; DGAPA-UNAM grants IG101320, IN111315, IN111716-3, IN111419, IA102019, IN112218; VIEP-BUAP; PIFI 2012, 2013, PROFOCIE 2014, 2015; the University of Wisconsin Alumni Research Foundation; the Institute of Geophysics, Planetary Physics, and Signatures at Los Alamos National Laboratory; Polish Science Centre grant, DEC-2017/27/B/ST9/02272; Coordinaci\'on de la Investigaci\'on Cient\'ifica de la Universidad Michoacana; Royal Society - Newton Advanced Fellowship 180385; Generalitat Valenciana, grant CIDEGENT/2018/034; Chulalongkorn University's CUniverse (CUAASC) grant; the European Union's Horizon 2020 research and innovation programme under the Marie Sk\l{}odowska-Curie grant agreement No 690575; Universidad de Costa Rica grants B9171, B6509, B5198. Thanks to Scott Delay, Luciano D\'iaz and Eduardo Murrieta for technical support.

\vspace{5mm}
\facilities{HAWC, \it{Fermi}}

\software{astropy \citep{2013A&A...558A..33A},  
          3ML \citep{threeML}, 
          naima \citep{naima},
          fermitools
          }


\bibliographystyle{aasjournal}
\bibliography{j2006}



\end{document}